\documentclass[%
reprint,
superscriptaddress,
showpacs,
amsmath,amssymb,
aps,
prl,
longbibliography,
]{revtex4-1}
\usepackage{psfrag,graphicx,epsfig,color}
\usepackage{dcolumn}
\usepackage{bm}
\usepackage{natbib}
\usepackage{float}
\usepackage[usenames,dvipsnames,svgnames,table]{xcolor}
\usepackage{subfigure}
\usepackage{nicefrac}

\graphicspath{{figs/}{plots/}}

\def\re    {{R_\lambda}}

\def\rr {{\mathbf{r}}}

\definecolor{mygreen}{rgb}{0,0.7,0.}

\begin{document}

\title{
Role of initial separation for Richardson scaling 
of turbulent two-particle dispersion
}

\author{Dhawal Buaria }
\email[]{dhawal.buaria@nyu.edu}
\affiliation{Tandon School of Engineering, New York University, New York, NY 11201, USA}
\affiliation{Max Planck Institute for Dynamics and Self-Organization, 37077 G\"ottingen, Germany}

\date{\today}

\begin{abstract}

The role of initial separation for inertial range scaling of 
turbulent two-particle dispersion is reassessed 
in light of recent results.
(slightly expanded version of Comment 
published in Physical
Review Letters \cite{buaria2023comment})

\end{abstract}

\maketitle


Using laboratory experiments and some
previously published data, 
Tan \& Ni \cite{tan_prl} investigate  
turbulent dispersion of particle-pairs,
characterized by the separation vector $\rr(t)$
as function of time $t$.
The authors report that the anticipated Richardson 
cubic scaling in the inertial-range:
\begin{align}
\langle |\rr(t) - \rr(0)|^2  \rangle = g \langle \epsilon \rangle t^3 \ , 
\end{align}
can only be realized at finite
(Taylor-scale) Reynolds number $\re$ for
one critical initial-separation of
$ r_0 = |\rr(0)| \approx 3\eta$,
$\eta$ being the Kolmogorov length-scale, 
or for any $r_0$ at infinite $\re$.
However, the authors' methodology to assess
Richardson scaling is unjustified, especially given
the $\re$ in their experiments is 
not high enough to produce Lagrangian inertial-range
characteristics. 
Indeed, when considering results
at significantly higher $\re=1000$ \cite{BSY.2015, BYS.2016} 
(currently the highest in Lagrangian
studies), which were ignored in \cite{tan_prl}, a different conclusion 
is reached.

\begin{figure}[h]
\begin{center}
\vspace{0.5cm}
\includegraphics[width=0.47\textwidth]{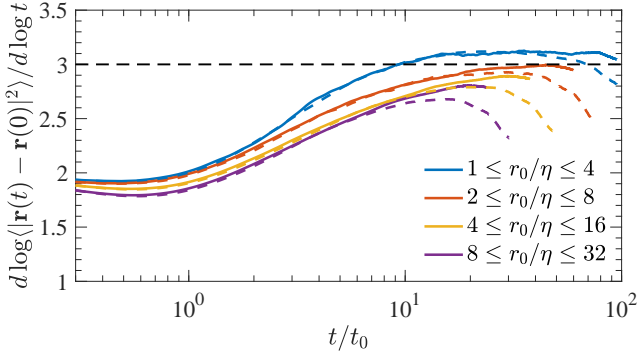}
\caption{
Log-local-slope of $\langle |\rr(t) - \rr(0)|^2 \rangle$
for various initial separations at $\re=1000$ (solid lines) and
$\re=650$ (dashed lines). 
} 
\label{fig:ls}
\end{center}
\end{figure}

By definition, Richardson scaling in the inertial-range 
requires $t_0 \ll t \ll T_L$, where
$t_0 = ( r_0^2/\langle \epsilon \rangle )^{1/3}$ is the Batchelor timescale,
characterizing the memory of initial ballistic separation
and $T_L$ is the Lagrangian integral time-scale, marking
the transition to diffusive behavior. 
However, Tan \& Ni \cite{tan_prl} 
assume a unique power-law 
$\langle |\rr(t) -\rr(0)|^2 \rangle \sim t^{k_D}$
in the range $2t_0 \le t \le T_L/2$ and thereby obtain
the exponents $k_D$ for various $r_0/\eta$. 
This choice of fitting-window is ad-hoc and
does not correctly reflect the inertial range behavior. 
In fact, a unique exponent cannot even be identified
in this range, as evident from
Fig.~\ref{fig:ls}, which shows the
log-local-slope of $\langle |\rr(t) - \rr(0)|^2 \rangle$
using the data in ref.~\cite{BSY.2015}.

\begin{table}
    \begin{tabular}{c|c|c|c}
\hline
\hline
    $r_0/\eta$ bin & effective $r_0/\eta$ & $k_D$ ($\re=1000$) & $k_D$  ($\re=650$) \\
\hline
    $1-4$    & 3  & 3.10  & 3.12   \\
    $2-8$    & 6  & 3.00  & 2.92  \\
    $4-16$   & 12 & 2.90  & 2.80  \\
    $8-32$   & 24 & 2.80  & 2.67  \\
\hline
    \end{tabular}
\caption{Inertial-range exponents $k_D$
as extracted from Fig.~\ref{fig:ls}.
}
\label{tab:kd}
\end{table}

Instead, we extract the exponents $k_D$
corresponding to where the curves
plateau in Fig.~\ref{fig:ls}. These exponents
are {\em uniquely} determined and also correspond to 
$t/t_0 > 10$, nominally satisfying
$t \gg t_0$ as required for inertial-range. 
The $k_D$ values are tabulated in Table~\ref{tab:kd}
for various initial-separations and are substantially
different than predicted by the proposed model in \cite{tan_prl}.
Evidently, Richardson scaling ($k_D=3$)
is obtained for $2< r_0/\eta < 8$
(with an effective $r_0/\eta\approx6$ 
\footnote{since the sample distribution is 
quadratic in $r_0$; see Fig.1 of
Ref.~\cite{BSY.2015}})
and trends with $\re$  suggest that 
an increasing range of $r_0/\eta$ will approach
$k_D=3$ at finite $\re$
(albeit significantly higher $\re$ than currently available 
are required to obtain a sufficient scale separation, 
as discussed later).

\begin{figure}[h]
\begin{center}
\includegraphics[width=0.41\textwidth]{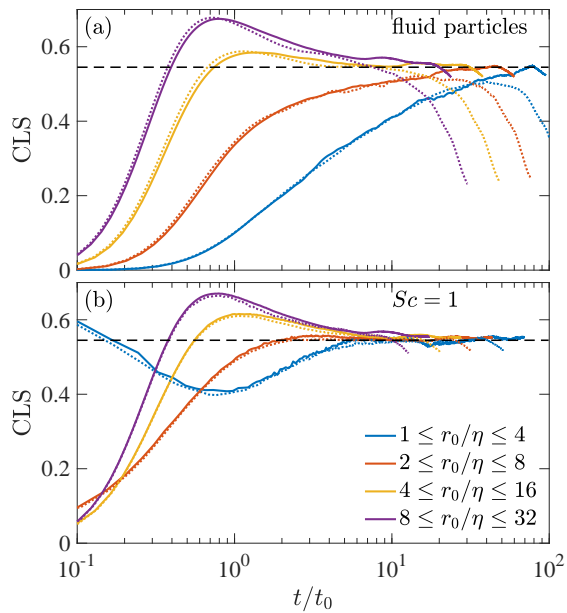}
\caption{
Cubed-local slope, as defined by Eq.~\eqref{eq:cls} 
for (a) fluid particles 
and (b) diffusive fluid particles with $Sc=1$,
for various initial separations.
Solid lines are for $\re=1000$ and dotted lines
for $\re=650$. Horizontal dashed line at $g=0.55$.
The legend applies to both panels.
}
\label{fig:cls}
\end{center}
\end{figure}

Concurrently, 
one can also consider the so-called
cubed-local-slope (CLS): 
\begin{align}
\frac{1}{\langle \epsilon \rangle} 
\left( \frac{d}{dt} [\langle |\rr(t)|^2\rangle ]^{1/3} \right)^3 \ , 
\label{eq:cls}
\end{align}
which should plateau at the same $g$ for various $r_0/\eta$ 
if the Richardson scaling holds.
Figure~\ref{fig:cls}a shows the CLS versus 
$t/t_0$ for various $r_0/\eta$, 
at $\re=1000$ and $650$
(reproduced from Fig.4a of \cite{BSY.2015}).
Evidently, based on the $\re$-trend, 
there is a clear tendency for various $r_0/\eta$
bins to approach $g\approx0.55$.
An even more robust result is obtained
in \cite{BYS.2016}, by considering 
particle-pairs which additionally undergo 
diffusive (Brownian) motion --
modeled by standard Wiener process --
enabling them to lose memory of their
initial separation faster.
Figure~\ref{eq:cls}b shows the CLS, 
corresponding to unity Schmidt number ($Sc=1$) 
for the diffusive motion
(reproduced from Fig.6 of \cite{BYS.2016}).
For this case, a very clear tendency to plateau at $g\approx0.55$ 
is observed for various $r_0/\eta$, for both $\re$.


Overall, these results indicate that Richardson scaling
is obtained in a critical range of $r_0/\eta$,
which slowly increases with $\re$, as opposed
to a single critical value of $r_0/\eta=3$.
It is well-known that the temporal scale-range in turbulence,
identified by $T_L/t_\eta$ 
($t_\eta$ being the Kolmogorov time-scale),
is already significantly smaller than the spatial
one, making robust inertial-range
observations of Lagrangian statistics very 
difficult \cite{sawford11}.
To observe Richardson
scaling, an additional constraint is imposed
by $t_0$, since $t_0 /t_\eta= (r_0/\eta)^{2/3}$
implying the range $T_L/t_0$ gets increasingly smaller
for $r_0/\eta>1$. At the highest $\re=1000$ currently 
available, $T_L/t_\eta\approx80$ \cite{BSY.2015},
and for say $r_0/\eta=8$, we have $t_0/t_\eta=4$, implying
$T_L/t_0 \approx 20$, which is barely one decade
of effective scale-separation. 
The scale separation for smaller $r_0/\eta$ would 
be larger. Considering the range
of $\re$ investigated by \cite{tan_prl} 
is noticeably lower than $\re=1000$, 
their conclusion 
that Richardson scaling holds for only 
$r_0/\eta\approx3$ can be readily explained
in terms of available scale-separation.
At higher $\re$, 
Richardson scaling would hold
for an increasing range of
$r_0/\eta$ -- as suggested by the trends
in Fig.~\ref{fig:ls} and Table~\ref{tab:kd}.


%

\end{document}